\documentclass[a4paper,fleqn]{cas-sc}

\usepackage[authoryear]{natbib}
\usepackage{lineno}
\def\tsc#1{\csdef{#1}{\textsc{\lowercase{#1}}\xspace}}
\tsc{WGM}
\tsc{QE}
\begin{document}
\let\WriteBookmarks\relax
\def\floatpagepagefraction{1}
\def\textpagefraction{.001}
%\linenumbers

% Short author
\shortauthors{P. Kumar et~al.}  

% Main title of the paper
\title [mode = title]{Investigation of an X-class Flare and Quasi-Periodic Pulsations in NOAA AR 13842 using SDO, Aditya-L1 and Solar Orbiter}  

\shorttitle{X-class Flare and QPPs in NOAA AR 13842} 

\author[1]{Pawan Kumar}[orcid=0009-0000-3324-2987]
\ead{pawankrchandrawanshi@gmail.com}
\credit{Conceptualization, Investigation, Visualization, Data curation, Software, Methodology, Writing-original draft}
\affiliation[1]{organization={Department of Physics, Patna University}, city={Patna}, postcode={800005}, state={Bihar}, country={Inida}}

\author[1]{Sanjay Kumar}[orcid=0000-0003-4578-6572]
\cormark[1]
\ead{sainisanjay35@gmail.com}
\credit{Conceptualization, Supervision, Funding acquisition, Writing-review and editing, Resources, Methodology}

\author[2]{Sushree S. Nayak}[orcid=0000-0002-4241-627X]
\ead{s.sangeetanayak93@gmail.com}
\credit{Conceptualization, Supervision, Writing-review and editing, Methodology}

\affiliation[2]{organization={Department de F\'isica, Universitat de les Illes Balears}, postcode={E-07122}, state={Palma de Mallorca}, country={Spain}}

\author[1]{Sadashiv}[orcid=0009-0003-0045-4387]
\ead{sadashyjh@gmail.com}
\credit{Visualization, Data curation, Investigation}

\author[3]{Sachin Rajkapoor Kanaujiya}[orcid=0009-0009-3048-0109]
\ead{sachin_p220261ph@nitc.ac.in}
\credit{Visualization, Data curation, Investigation, Writing}

\affiliation[3]{organization={Department of Physics, National Institute of Technology Calicut}, city={Kozhikode}, postcode={673601}, state={Kerala}, country={India}}

\author[4]{Ramesh Chandra}[orcid=0000-0002-3518-5856]
\ead{rchandra.ntl@gmail.com}
\credit{Funding acquisition, Formal analysis, Validation}

\affiliation[4]{organization={Department of Physics, DSB Campus, Kumaun University}, city={Nainital}, postcode={263001}, state={Uttrakhand}, country={India}}

\cortext[1]{Corresponding author. Email: sainisanjay35@gmail.com}

\begin{abstract}
We investigate an X9.0 class flare (SOL2024-10-03T12:08) and associated quasi-periodic pulsations (QPPs) in NOAA active region (AR) 13842 using observations from the Solar Dynamics Observatory (SDO), Aditya-L1 and Solar Orbiter (SO). To explore the role of magnetic configuration in the flare, we employ non-force-free-field extrapolation technique for reconstructing a sequence of the coronal magnetic field throughout the flare. The observations reveal the emergence of intricate flare brightenings and ribbons. QPPs are observed with periods of $\approx$ 2 and $\approx$ 6 minutes in EUV, $\approx$ 7 minutes in UV, and $\approx$ 1 minute in X-rays. The extrapolation documents the presence of a flux rope, a sheared arcade, and a hyperbolic flux tube (HFT) enveloping the rope and the arcade. The sequence of the extrapolation illustrates an increase in the twist of the rope as it ascends towards a torus-unstable region, thereby rendering the rope a viable candidate for eruption. The sheared arcade appears to generate electric current through its foot-point's motion, which explains the sigmoidal brightening observed at the flare onset. The motion also seems to drive the non-parallel field lines closer together, heightening gradient in the magnetic field beneath the rope. Consequently, recurrent magnetic reconnection is anticipated within the sheared arcade --- explaining the ascent of the rope with its twist increment. The formation of parallel flare ribbons is attributed to these reconnections. Furthermore, reconnection at the HFT is proposed to lead the remote and quasi-circular ribbons. Crucially, repeated reconnections inside the arcade as well as at the HFT are speculated to cause the QPPs.
\end{abstract}

\begin{keywords}
Solar flare \sep Extrapolation \sep Magnetic reconnection \sep Hyperbolic flux tube \sep Quasi-Periodic Pulsations
\end{keywords}

\maketitle
\section{Introduction}
\label{intro}

Solar flares represent highly energetic events occurring within the solar atmosphere. These events signify the abrupt and explosive release of a significant amount of magnetic energy from the solar corona \citep{2011LRSP....8....6S}. It is believed that magnetic reconnection, a process that involves a change in magnetic topology accompanied by a rapid conversion of magnetic energy into mass flow, heat, and the acceleration of charged particles, plays a vital role in the generation of flaring events \citep{2023AdSpR..71.1856P}. However, the physics governing magnetic reconnection, particularly in the complex magnetic configurations found in the flaring regions, is not yet fully comprehended \citep{2011LRSP....8....6S,2023AdSpR..71.1856P}.

In a three-dimensional (3D) magnetic configuration, quasi-separatrix layers (QSLs) are areas characterized by significant gradients in magnetic field line connectivity, making them potential locations for reconnection \citep{1999A&A...351..707T,2002JGRA..107.1164T,2006SoPh..238..347A}.
QSLs are often associated with increased electric current densities \citep{2009ApJ...700..559M,2009ApJ...704.1288W} and demonstrate a significant relationship with the observed flare ribbons \citep{1997A&A...325..305D,2013A&A...555A..77J}. Magnetic reconnection occurring within QSLs is generally linked to the apparent slipping motion of magnetic field lines, which results from the ongoing reconnection-driven exchange of magnetic connectivity \citep{1995JGR...10023443P,2003JGRA..108.1285P,2006SoPh..238..347A,2012A&A...543A.110A,2013A&A...555A..77J}. Such reconnection is considered a key mechanism for initiating solar flares \citep{2006SoPh..238..347A,2023A&A...677A..43P,2025SoPh..300...22K}. Further, the intersection of two QSLs forms a hyperbolic flux tube (HFT), a 3D magnetic structure characterized by a pronounced concentration of magnetic connectivity gradients \citep{1999A&A...351..707T,2002JGRA..107.1164T}. The HFT provides a preferred site for the formation of intense current layers and magnetic reconnection, and reconnection within or around the HFT has been suggested to play an important role in the initiation and evolution of solar flares \citep{2017A&A...604A..76M,2021MNRAS.503.1017M,2024SoPh..299...15A,2026SoPh..301...67K}.

Notably, eruptive flares are commonly associated with magnetic flux ropes --- magnetic structures consist of twisted magnetic field lines that wrap around a central axis \citep{2021PhPl...28b4502N,2022FrASS...939061K}. Observing flux ropes directly is difficult due to their relatively low plasma density; nevertheless, prominence and filament eruptions offer compelling evidence that twisted magnetic structures can be present in the corona before the onset of flares \citep{2012A&A...540A.127K}. Furthermore, coronal sigmoids, which are defined as S-shaped or inverse S-shaped loop systems seen in soft X-ray (SXR) and extreme ultra-violet (EUV) wavelengths \citep{1996ApJ...464L.199R}, are frequently interpreted as indicators of flux rope formation \citep{2009ApJ...700L..83G,2011A&A...526A...2G,2012ApJ...759..105S}.

An interesting phenomenon associated with the flare emission is quasi-periodic pulsations (QPPs). QPPs are often characterized by a number of repetitive, impulsive, and successive pulsations observed in time dependent intensity curves during solar flares \citep{2009SSRv..149..119N}. Specifically, QPP denotes a flare time series that includes at least three or more consecutive pulsations \citep{2018SSRv..214...45M,2021ApJ...910..123C,2024A&A...684A.215C,2024ApJ...971...29I}. However, the observed QPPs often have varying instantaneous periods which are interpreted as ``quasi-periods'' \citep{2019PPCF...61a4024N}. QPPs were first identified and noted by \citet{1969ApJ...155L.117P} in the microwave and hard X-ray (HXR) wavebands. Later, they are observed in all electromagnetic wavelengths, such as gamma ($\gamma$)- rays, SXR, HXR, ultraviolet (UV), EUV, H-alpha (H$\alpha$), Lyman-alpha (Ly$\alpha$), white light, and radio/microwave \citep{2001ApJ...562L.103A,2005ESASP.596E..46F,2006ApJ...644L.149O,2008SoPh..247...77L,2008MNRAS.388.1899S,2009SSRv..149..119N,2011A&A...533A..61G, 2011A&A...533A.116Y,2016ApJ...819L..24Y,2016ApJ...827L..30H,2019ApJ...875...33H,2020STP.....6a...3K,2025A&A...695A.136B,2026ApJ..1003...78K}. Various physical mechanisms such as the periodic occurrence of magnetic reconnection, the interaction of magnetohydrodynamic (MHD) waves with the reconnection site are proposed to be responsible for QPPs \citep{2010PPCF...52l4009N,2016ApJ...822....7K,2021SSRv..217...66Z,2026NatAs..10...54A,2026NatAs..10...13L}. However, the origin of QPPs is not yet fully understood and requires further investigation \citep{2016SoPh..291.3143V,2018SSRv..214...45M}.

In this paper, we investigate the onset of an X9.0 class flare having a complex morphology of flare ribbons and associated QPPs, hosted by NOAA active region (AR) 13842 on 2024 October 3. In a recent study, \citet{2025ApJ...985L..16D} suggested that the collisional shearing is responsible for flux rope formation and triggering this flare using multi-wavelength observations and the non-linear force free field (NLFFF) extrapolation modeling.  
\citet{2025ApJ...991...77M} performed a data constrained MHD simulation initiated by the NLFFF to simulate the flare dynamics. They proposed that the tether cutting magnetic reconnection in sheared field lines led to flux rope eruption and formation of the flare ribbons. Additionally, \citet{2025A&A...695L...4L} investigated the short period pulsations in $\gamma$-rays and the higher energy range ($\geq20$ keV) of HXR during this flare and reported QPPs at a shorter period of about 1 second. Here we find QPPs with multiple periods of $\approx$ 2 minutes and $\approx$ 6 minutes in EUV, $\approx$ 7 minutes in UV channels, and $\approx$ 1 minute in HXR wavebands. Furthermore, we consider a sequence of non-force-free-field (NFFF) extrapolations \citep{2010JASTP..72..219H} for a time period ranging from 11:00 UT to 12:24 UT to examine the evolution of magnetic configuration in the flaring region and its role in producing the flare brightenings and the ribbons. Importantly, the extrapolated magnetic configuration documents the presence of a HFT and a sheared arcade enveloping a flux rope in the flaring region. Recurrent magnetic reconnections at these magnetic features are proposed to play a central role in producing the flare and associated QPPs.

This paper is arranged as follows, Section \ref{Obs} briefly discusses the key multi-wavelength observations of the flare and analysis of associated QPPs in different passbands. Section \ref{exp-mod} documents the NFFF extrapolations of AR 13842. Section \ref{result} elaborates on the key magnetic features in the flaring region and their evolution through the series of extrapolations, along with the relation of evolution to the observation of the flare. Lastly, we present a brief summary and key findings of the paper in Section \ref{Summary}.

\section{Observation of the X-Class flare in AR 13842 and associated QPPs}
\label{Obs}

\subsection{Flare history from AIA/SDO}
\vspace{0.2 cm}

\begin{figure}
   \centering
    \includegraphics[width=1.0\linewidth]{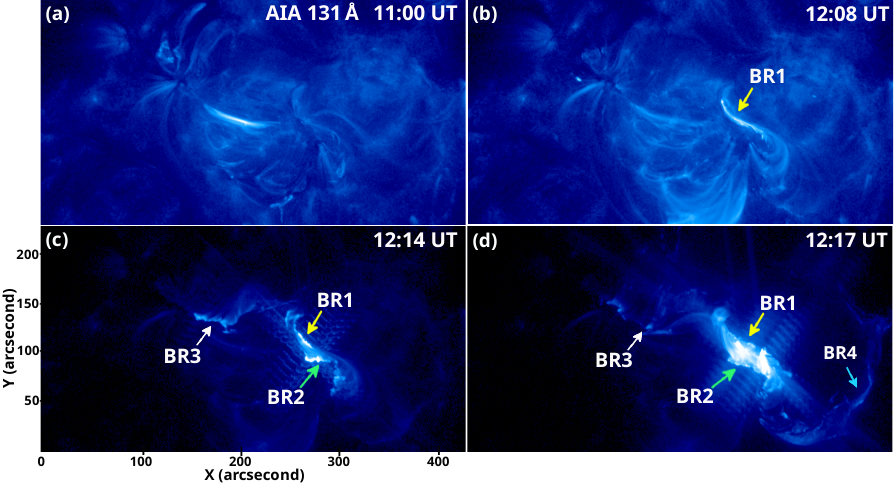}
    \caption{The time evolution of the flare in AIA 131 {\AA} channel. Panel (a) show pre-flare stage at $\approx$ 11:00 UT. Panel (b) highlight the appearance of an S-shaped brightening BR1 (yellow arrow) at the flare onset ($\approx$ 12:08 UT) near the flare kernel. Later, the appearance of another flare brightening BR2 (green arrow) and a remote brightening BR3 (white arrow) are displayed in panel (c) at $\approx$ 12:14 UT. Close to flare peak time, an additional quasi-circular brightening BR4 (cyan arrow) is found to appear at $\approx$ 12:17 UT (panel (d)). (An animation of this figure is available.)}
     \label{131_obs}
\end{figure}

\begin{figure}
   \centering
    \includegraphics[width=1.0\linewidth]{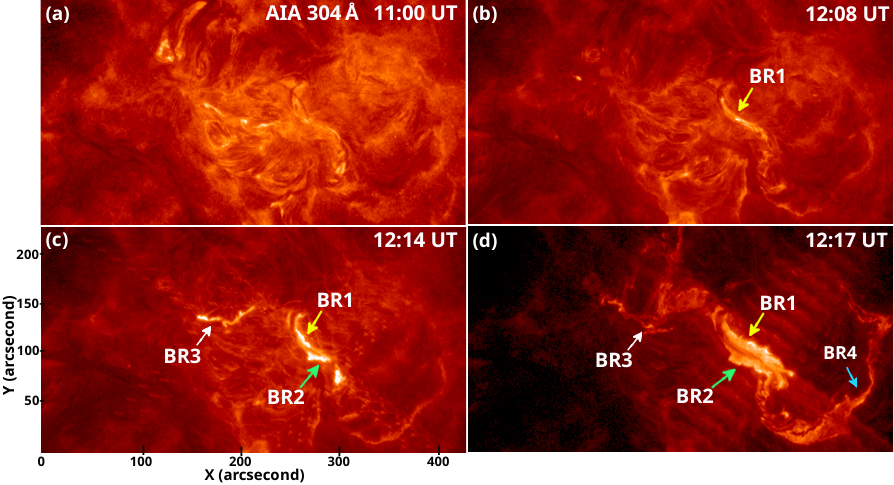}
    \caption{Temporal progression of the flare brightenings in the AIA 304 {\AA} channel. Panel (a) illustrates the pre-flare phase at approximately 11:00 UT. In panel (b), the emergence of an S-shaped brightening, designated as BR1 (indicated by the yellow arrow), is noted at the onset of the flare around 12:08 UT, in the flaring region. Subsequently, panel (c) reveals the emergence of another flare brightening, BR2 (marked by the green arrow), along with a remote brightening, BR3 (shown by the white arrow), at approximately 12:14 UT. Near the peak of the flare, an additional quasi-circular brightening, BR4 (identified by the cyan arrow), is observed at around 12:17 UT in panel (d). (An animation of this figure is available.)}
     \label{304_obs}
\end{figure}

The X9.0 class flare occurred on October 3, 2024 and hosted by NOAA AR 13842. The flare started at 12:08 UT, peaked at 12:18 UT and ended at 12:27 UT \citep{2025ApJ...991...77M}. To explore the evolution of flare brightenings and the ribbons, we have employed the multi-wavelength images supplied by the Atmospheric Imaging Assembly \citep[AIA,][]{2012SoPh..275...17L} onboard the Solar Dynamics Observatory \citep[SDO,][]{2012SoPh..275....3P} with a time resolution of 12 seconds in EUV and 24 seconds in UV channels. Figure \ref{131_obs} presents the flare evolution in AIA 131 {\AA} starting from 11:00 UT. Panel (a) corresponds to observations at a pre-flaring time. Around flare onset time ($\approx$ 12:08 UT), we observe an S-shaped brightening, marked as BR1 with a yellow arrow in panel (b). Subsequently, BR1 grows in time along with appearance of another brightening BR2 opposite and close to BR1 (denoted by a green arrow in panel (c)). Moreover, the panel documents generation of a remote brightening BR3 highlighted by a white arrow. As time advances toward the flare peak, a prominent quasi-circular shaped brightening BR4 is also seen in the flaring region (cyan arrow in panel (d)). Notably, this quasi-circular brightening is also observed in $\text{H} \alpha$ and $\text{Ly} \alpha$ images \citep{2025A&A...695L...4L}. Next, Figure \ref{304_obs} illustrates the flare development in AIA 304 {\AA}. Interestingly, the figure also advocates the appearance of four flare brightenings akin to 131 {\AA}. It is noteworthy that the BR4 exhibits a brighter appearance in 304 {\AA}, indicating that the emission is primarily dominated by the lower transition region and upper chromosphere, with its effects extending into the coronal region. 

\begin{figure}
	\centering
	\includegraphics[width=1.0\linewidth]{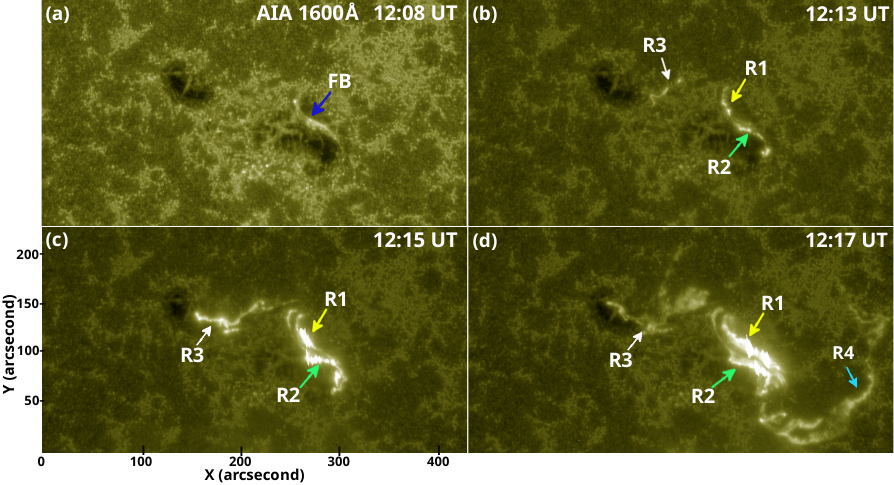}
	\caption{Formation of flare ribbons in AIA 1600 {\AA}. Panel (a) shows initial faint brightening (FB, blue arrow) around the flare onset time. Panel (b) depicts further growth of the ribbons from the flare kernel marked as R1 (yellow arrow) and R2 (green arrow) along with a distant ribbon R3 (white arrow) at $\approx$ 12:13 UT. Panel (c) shows the development of R1, R2 and R3 ($\approx$ 12:15 UT). Panel (d) presents another faint quasi-circular ribbon R4 (cyan arrow), fully appeared at $\approx$ 12:17 UT. Also notable is the expansion of R1 and R2 --- making them increasingly parallel to each-other. (An animation of this figure is available.)}
    \label{1600_obs}
\end{figure}

Figure \ref{1600_obs} presents observations of the flare in AIA 1600 {\AA}. In panel (a), a faint S-shaped brightening FB, similar to the ones observed in Figures \ref{131_obs}(b) and \ref{304_obs}(b), is seen (pointed by a blue arrow) --- marking the onset of the flare. Later, we observe the appearance of three flare ribbons in which two ribbons R1 (yellow arrow) and R2 (green arrow) are in close proximity, and the ribbon R3 (white arrow) is located away from R1 and R2 (see panel (b)). With time, the ribbons expand and the separation between R1 and R2 also becomes apparent (panel (c)). Near the flare peak time, the ribbons R1 and R2 become nearly parallel to each-other (panel (d)). Simultaneously, another quasi-circular flare ribbon R4 is developed, pointed by a cyan arrow in panel (d). Notably, from Figures \ref{131_obs}, \ref{304_obs} and \ref{1600_obs}, the brightenings BR1-BR4 and the ribbons R1-R4 are found to have a similar morphology.  

\subsection{QPP analysis from AIA/SDO}
\vspace{0.2 cm}

\begin{figure}
   \centering
    \includegraphics[width=1.0\linewidth]{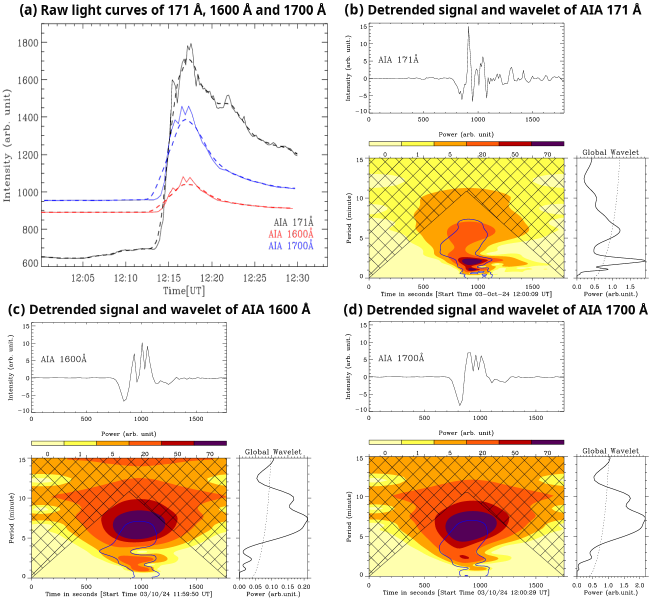}
    \caption{Panel (a) plots the raw light curves of AIA 171 {\AA} (black curve), 1600 {\AA} (blue curve) and 1700 {\AA} (red curve) starting from 12:00 UT to 12:30 UT. The panel is further superimposed with the smooth background (dashed lines). Panel (b) presents the detrended light curves (at top of panel) and respective Morlet wavelet (in bottom left) and global wavelet spectra (in bottom right) in 171 {\AA}. The blue contour overlaid with the Morlet wavelet encompasses the statistically significant power exceeding the 95 {\%} confidence level. Also, global wavelet spectra reveals the oscillation periods crossing the 95$\%$ red-noise (dashed curve). The analysis reveals the two statistically significant oscillation periods in 171 {\AA} approximately $2.13 \pm 0.29$ minutes and $6.00 \pm 0.29$ minutes in AIA 171 {\AA}. Panels (c) and (d) demonstrate the similar analysis for 1600 {\AA} and 1700 {\AA}, indicating the presence of statistically significant periods $\approx$ $7.40 \pm 0.58$ minutes and $\approx$ $7.15 \pm 0.58$ minutes, respectively.} 
     \label{qpp_aia}
\end{figure}

To identify the QPPs associated with flare emission, we have analyzed the wavelet spectra of UV and EUV channels of AIA/SDO. To achieve this, we first examine the time evolution of the flare emission in the UV and EUV passbands from 12:00 UT to 12:30 UT, which encompasses the entire duration of the flare. During this selected time frame, several UV and EUV channels are highly affected by the saturation around the flare peak. Consequently, only AIA 171 {\AA} (in EUV), 1600 {\AA} and 1700 {\AA} (in UV) observations are considered for the further analysis. In Figure \ref{qpp_aia}, panel (a) illustrates the raw light curves (solid lines) of the AIA 171 {\AA} (black curve), 1600 {\AA} (red curve) and 1700 {\AA} (blue curve) extracted from their respective intensity maps of the flaring region. It is important to note that the AIA exposure times vary throughout the impulsive phase of the flare. Instead of limiting the dataset to shorter-exposure images, we used all available observations at the native AIA cadence to maintain continuous temporal coverage of the event. To minimize the impact of saturated pixels, the light curves are constructed by averaging only those pixels within the designated flaring area that maintained intensities below the saturation threshold established at the peak of the flare. Notably, light curves exhibit three prominent intensity enhancements during the impulsive phase of the flare in all three passbands. The panel is further superimposed with the smoothed background trends (dashed lines), obtained by applying a smoothing filter \citep{Peng1999} with a window size of 8 data points on the respective raw light curves. Subsequently, the smooth trend is subtracted from the original intensity profile to obtain the detrended signals, which are suitable for periodicity analysis in QPP studies. Furthermore, we have used the detrended signals to perform the wavelet transform using the Morlet wavelet method \citep{1998BAMS...79...61T} for all wavebands. 

In Figures \ref{qpp_aia}(b-d), the detrended light curves in the EUV and UV wavebands are displayed at the top of each panel, accompanied by their respective wavelet spectra at the bottom left of each panel and the global wavelet spectra at the bottom right of each panel. Also, the blue contours overlaid with the Morlet wavelet spectra delineates regions of statistically significant power that surpass the 95$\%$ red-noise significance level, whereas the cross-hatched areas indicate the cone of influence where edge effects reduce the reliability of the derived periodicities. Additionally, the global wavelet spectra are used to identify the dominant oscillation periods crossing the 95$\%$ red-noise significance level (dashed curve in global wavelet plot) in each wavelength. The wavelet analysis reveals two statistically significant periodic components with characteristic periods of approximately $2.13 \pm 0.29$ minutes and $6.00 \pm 0.29$ minutes in AIA 171 {\AA} (Figure \ref{qpp_aia}(b)), indicating the presence of multi-periodic QPPs \citep{2019ApJ...878...78C,2025JGRA..13033772L}. In contrast, the UV channels are dominated by longer-period oscillations of approximately $7.40 \pm 0.58$ minutes in 1600 {\AA} (Figure \ref{qpp_aia}(c)), and $7.15 \pm 0.58$ minutes in 1700 {\AA} (Figure \ref{qpp_aia}(d)). The presence of different characteristic periods in the EUV and UV channels indicates that the flare exhibits multi-periodic intensity variations, with distinct temporal behavior observed at different atmospheric heights sampled by these passbands.

\subsection{QPP analysis from HEL1OS/Aditya-L1}
\label{EEMD_hel1os}
\begin{figure}
   \centering
    \includegraphics[width=1.0\linewidth]{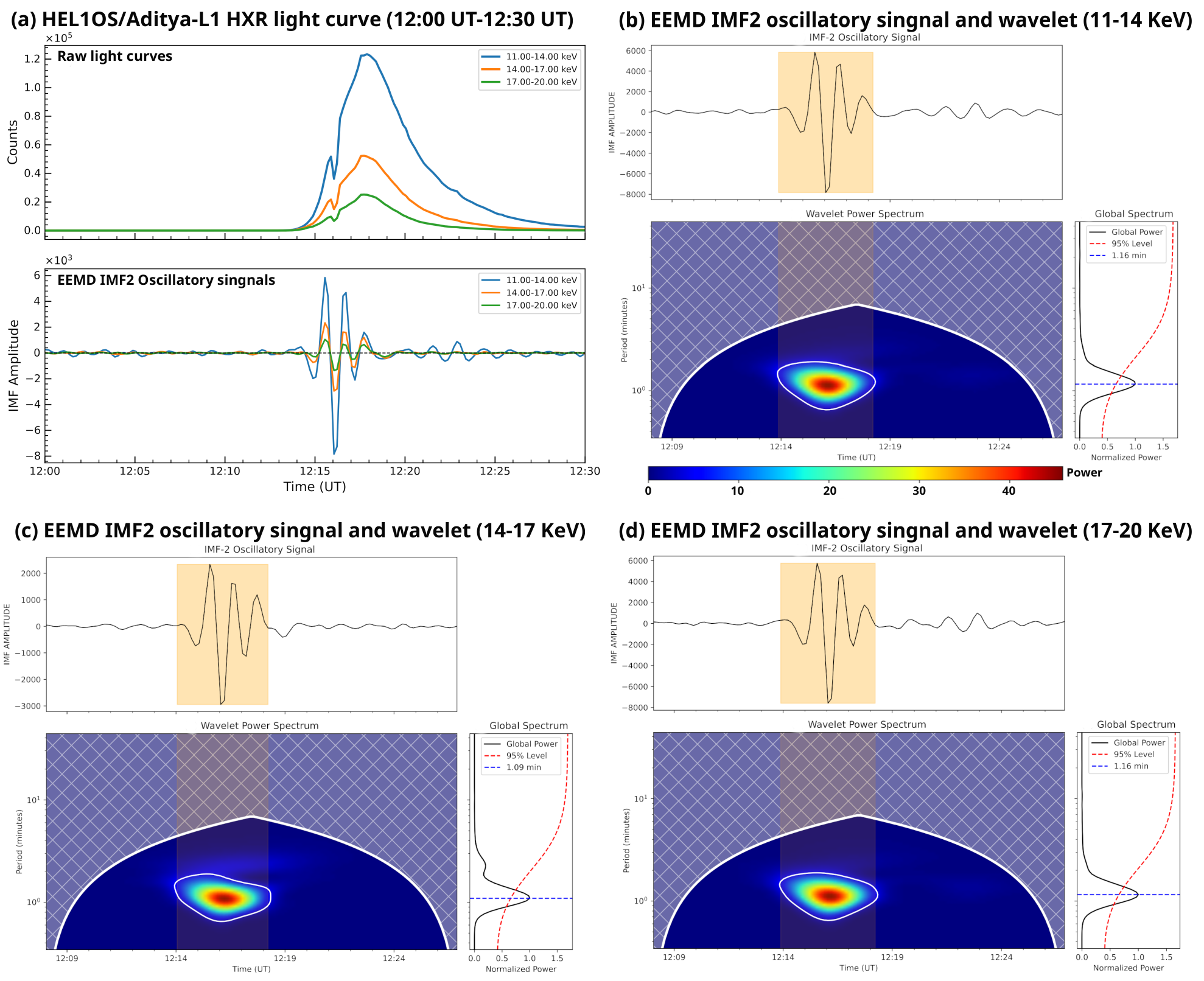}
    \caption{Top of panel (a) shows the energy resolved HXR raw light curves in 11--14 keV (blue curve), 14--17 keV (yellow curve), and 17--20 keV (green curve) energy bands. The bottom of panel (a) presents the corresponding IMF-2 derived from EEMD method. Panel (b) shows the IMF-2 signal (top) and respective Morlet wavelet (bottom left) and global wavelet spectra (bottom right) in 11--14 keV. The white contour overlaid with the Morlet wavelet encompasses the statistically significant power exceeding the 95$\%$ confidence level. Also, global wavelet spectra reveals the oscillation periods (blue dashed line) crossing the 95$\%$ red-noise (red curve). This reveals statistically significant oscillation period of approximately $1.16 \pm 0.22$ minutes in the 11--14 keV. The similar investigations are done for 14--17 keV and 17--20 keV energy bands in panels (c) and (d) respectively, which indicates the periods of significant oscillations about $1.09 \pm 0.22$ minutes in the 14--17 keV and $1.16 \pm 0.22$ minutes in the 17--20 keV.  }
     \label{qpp_hel1os}
\end{figure}

\vspace{0.2 cm}
To explore the presence of QPPs signatures in HXR emission of the flare, we have utilized the observations from the High Energy L1 Orbiting X-ray Spectrometer \citep[HEL1OS,][]{2025SoPh..300..140N} onboard Aditya-L1 \citep{2025SoPh..300..128P}. Notably, the original light curve obtained from HEL1OS is rebinned into 10 second time bins to improve the signal-to-noise ratio and suppress small-timescale statistical fluctuations. Panel (a) of Figure \ref{qpp_hel1os} presents the energy resolved HXR raw light curves in 11--14 keV (blue curve), 14--17 keV (yellow curve), and 17--20 keV (green curve) energy bands from 12:00 UT to 12:30 UT. Notably, all three energy bands demonstrate a prominent peak in the HXR flux during the impulsive phase of the flare. Furthermore, to extract the oscillatory signals from the slowly varying flare background, we employ the Ensemble Empirical Mode Decomposition (EEMD) method \citep{1998RSPSA.454..903H,Wu2009EnsembleEM} with 10,000 ensemble trials and a noise width of 0.2 similar to \citet{2026ApJ..1003...78K}. This returns several intrinsic mode functions (IMFs), representing the fluctuations at different temporal scales (see Appendix \ref{appendix1}). The statistical significance of the derived IMFs is examined using Fourier and Morlet wavelet analysis. The analysis suggests that, among the extracted IMFs, only the second intrinsic mode function (IMF-2) has a statistically significant oscillatory power across aforementioned energy bands, leading to its selection for further investigation. The corresponding IMF-2 oscillatory signals are plotted in the bottom of Figure \ref{qpp_hel1os}(a).

\begin{figure}
   \centering
    \includegraphics[width=1.0\linewidth]{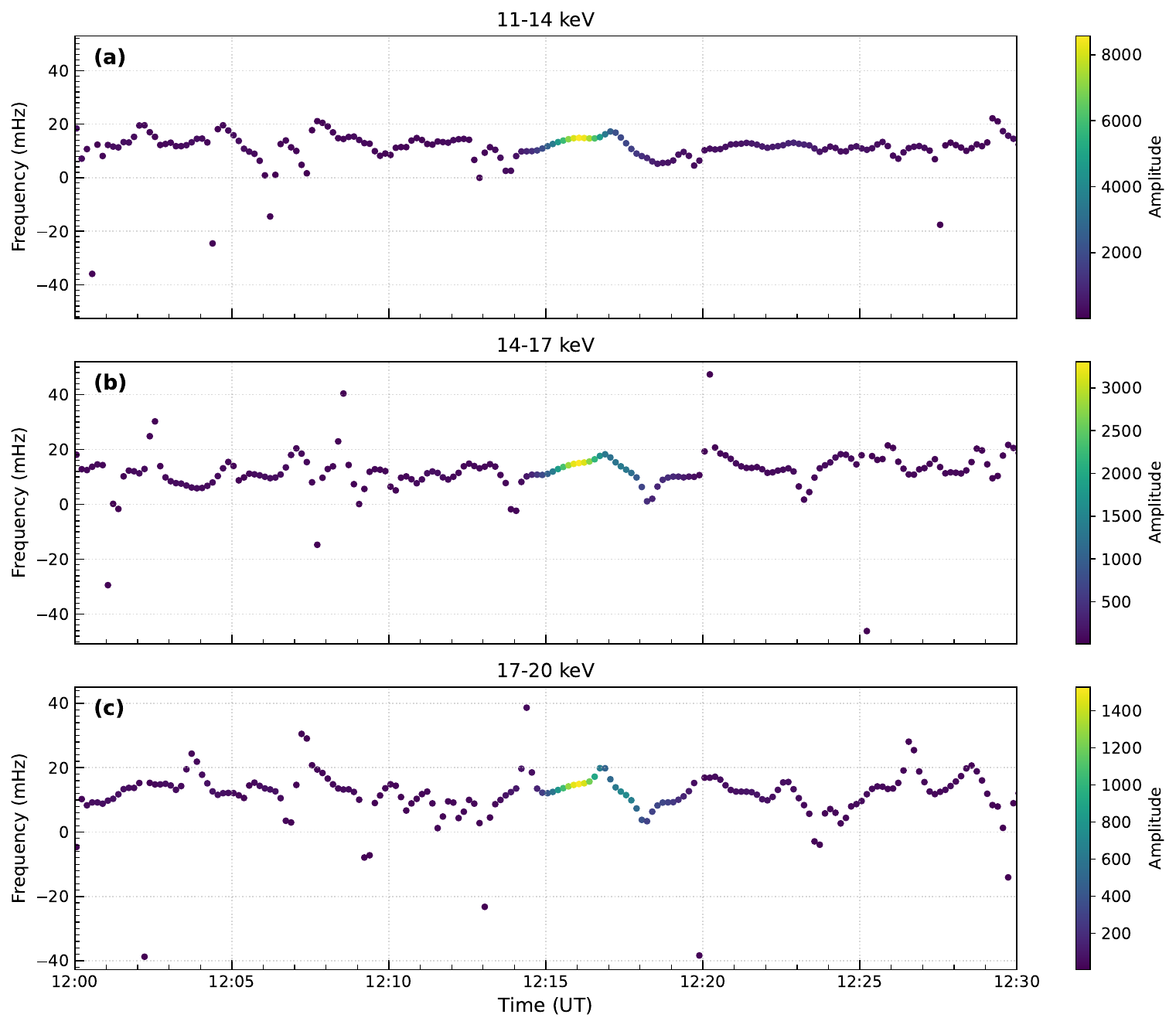}
    \caption{The Hilbert spectra obtained from the Hilbert transform using the IMF-2 signals. Panels (a), (b), and (c) correspond to energy bands 11--14 keV, 14--17 keV, and 17--20 keV, respectively. The spectra reveals that the oscillation is concentrated around $\approx$ 13 mHz in all energy bands.  }
     \label{qpp_hilbert}
\end{figure}

Figures \ref{qpp_hel1os}((b)-(d)) plot the selected IMF-2 signals (top of the panels) with their respective Morlet wavelet spectra (left bottom of the panels) and global wavelet spectra (right bottom of the panels) in 11--14 keV, 14--17 keV and 17--20 keV energy bands, respectively. The shaded region of the oscillatory signal highlights the time frame over which the oscillation satisfies the adopted significance criteria. The white contours overlaid with the wavelet spectrum signify the statistical power exceeding the 95$\%$ red-noise level. Also, the global wavelet spectra illustrates the periods (blue dashed lines) of oscillations crossing the 95$\%$ red-noise (red dashed line) confidence level. The overall wavelet analysis reveals the presence of QPPs signatures in all three energy bands with dominant periods approximately $1.09 \pm 0.22$ minutes in the 14--17 keV, alongside approximately $1.16 \pm 0.22$ minutes in both the 11--14 keV and 17--20 keV. Next, to investigate the temporal evolution of the oscillations, we have utilized the Hilbert-Huang transform \citep{1998RSPSA.454..903H} to analyze the selected IMF-2 signals. The obtained Hilbert spectra are shown in Figure~\ref{qpp_hilbert}, where panels (a)--(c) correspond to the 11--14 keV, 14--17 keV, and 17--20 keV energy bands, respectively. The color scale represents the instantaneous oscillation amplitude, while the vertical axes illustrate the instantaneous frequencies. The amplitude-weighted mean instantaneous frequencies are determined to be 13.23 mHz, 13.26 mHz, and 13.51 mHz for the 11--14 keV, 14--17 keV, and 17--20 keV channels, respectively. These values correspond to amplitude-weighted mean periods of $75.56 \pm 13.05$ seconds, $75.39 \pm 13.05$ seconds, and $73.99 \pm 13.05$ seconds for the respective energy bands. Notably, under the the estimated uncertainties, the Hilbert-derived periods are consistent across all three energy bands and in agreement with those obtained from the wavelet analysis. The oscillatory power remains concentrated at $\approx$ 13 mHz throughout the significant interval in considered energy bands, indicating minor temporal variations in the instantaneous frequency. Therefore, we can assert that the HEL1OS observations offer consistent evidence for QPPs within the 11--20 keV energy range.

\subsection{QPP analysis from STIX/Solar Orbiter}
\vspace{0.2 cm}

\begin{figure}
    \centering
    \includegraphics[width=1.0\linewidth]{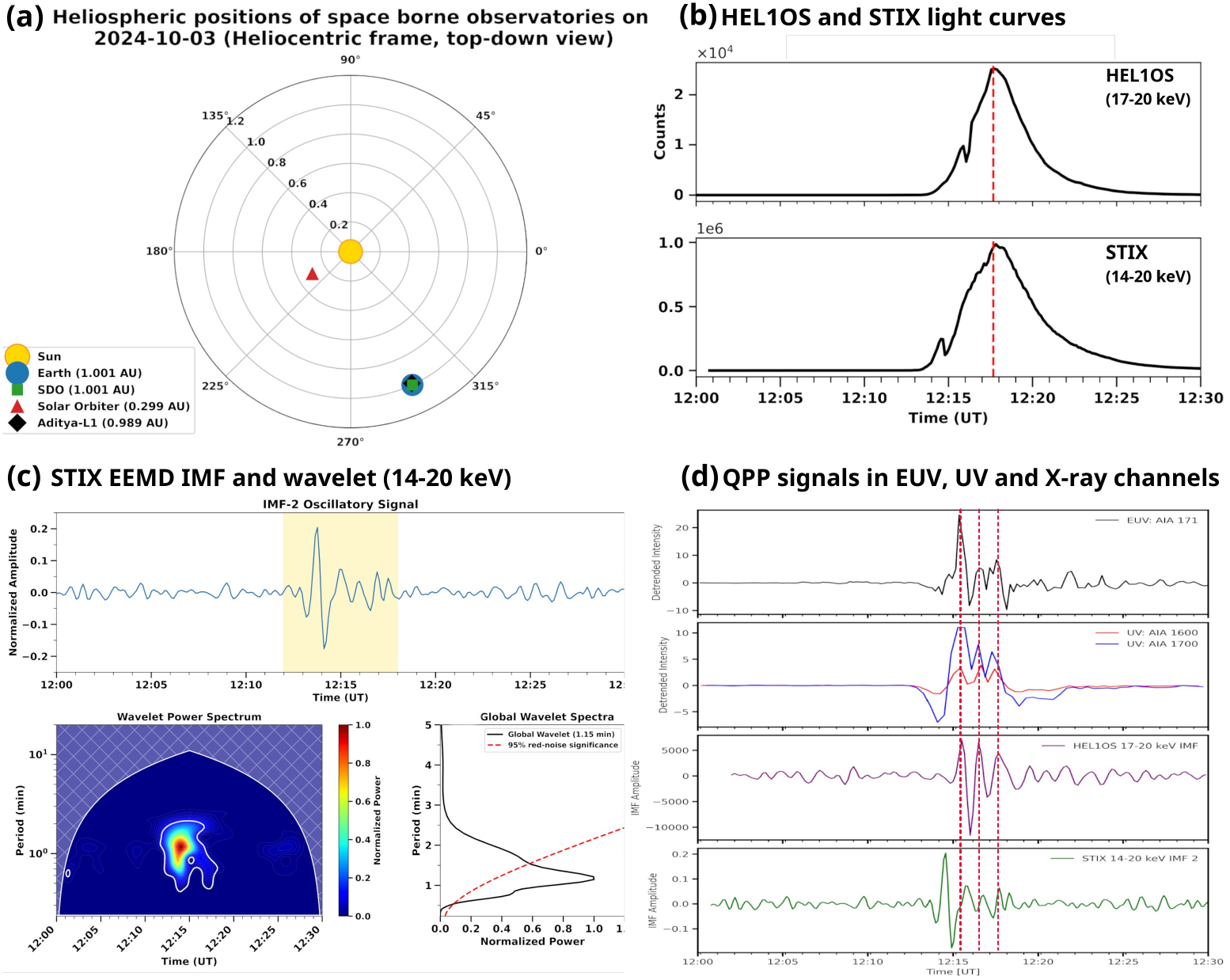}
    \caption{Panel (a) illustrates the spacecraft configuration in the heliocentric inertial frame. Panel (b) shows co-temporal X-ray light curve peaks (vertical red lines) for HEL1OS (top) and STIX (bottom). Panel (c) displays the STIX EEMD IMF-2 signal alongside its Morlet wavelet spectrum, where white contour encompasses power exceeding the 95$\%$ confidence level, alongside the global wavelet spectrum, which identifies an oscillation period of $\approx 1.15 \pm 0.3$ minutes crossing 95$\%$ red-noise significance (red dashed curve). Panel (d) compares the EUV (black), UV (red, blue), and HXR (purple for HEL1OS, green for STIX) input QPP signals. In the panel, the vertical dashed lines indicate the QPP peak times identified from the HEL1OS IMF profile and are overplotted to facilitate comparison across different wavebands.}  
    \label{qpp_stix}
\end{figure}

The Spectrometer/Telescope for Imaging X-rays \citep[STIX,][]{2020A&A...642A..15K} onboard the Solar Orbiter \citep[SO,][]{2020A&A...642A...1M}, records the X-ray data from the visible solar disk, also constructs the images of X-ray sources in the solar atmosphere with an angular resolution of $7''$. Because of the unique orbit of the SO, the STIX observes the flare from different vintage point with respect to the SDO and Aditya-L1 as shown in panel (a) of Figure \ref{qpp_stix}. The distance between the SO and the Sun was $\approx 0.299\text{ a.u.}$ at the time of flare. To facilitate a qualitative comparison with the SDO and Aditya-L1 observations, light-travel-time corrections were applied to the STIX observations prior to analysis, accounting for the observational time offset of approximately 344 seconds.
 Here we have utilized the STIX spectrogram data for the QPP investigation. This data product is summed over all imaging detectors and pixels, retaining full spectral resolution but no spatial information \citep{2020A&A...642A..15K}. To improve the signal-to-noise ratio while retaining sensitivity to the QPP timescales of interest, the data were rebinned to a 7 second cadence in the 14–20 keV energy range. The resulting light curve was then used for the subsequent QPP analysis. 

Figure \ref{qpp_stix}(b) displays the light curves from HEL1OS (top) and STIX (bottom), demonstrating co-temporal X-ray peaks (marked by vertical red lines) in both instruments. Next, to isolate the oscillatory signals from the background trend in the STIX light curve, the EEMD technique is employed akin to HEL1OS (see subsection \ref{EEMD_hel1os} and appendix \ref{appendix1}). Among the resulting EEMD IMFs, IMF-2 is  selected as the suitable QPP candidate in the 14–20 keV energy range and is therefore used as the input signal for the subsequent wavelet analysis. Panel (c) of the figure plots the selected IMF-2 signal (at top of panel) with respective Morlet wavelet power (bottom left of panel) and global wavelet spectra (bottom right of panel). The wavelet spectrum reveals that the maximum power exceeding the significance of 95$\%$ red-noise (white contour) is concentrated at around 12:15 UT, alongside the corresponding peak in the global wavelet spectra that crosses the 95$\%$ confidence level (red curve) with a period $\approx$ 1.15 minutes. Hence, the analysis suggests the presence of QPP with period approximately $1.15 \pm 0.3$ minutes in 14-20 keV STIX X-ray time series. Panel (d) presents the input QPP signals in the EUV (black curve), UV (blue and red curves), and HXR (purple curve for HEL1OS and green curve for STIX) channels. The vertical red dashed lines mark the peak times of the QPPs observed by HEL1OS and are superimposed on the other channels to facilitate comparison across different wavebands. The signals exhibit nearly co-temporal oscillations in the EUV, UV, and HXR emissions, suggesting a close temporal correspondence between the QPPs observed in these spectral ranges. The individual peaks show small temporal offsets of approximately 5–35 seconds between the channels. These offsets may arise from differences in the temporal resolution of the instruments and the distinct physical processes governing emission at different wavelengths. Nevertheless, the observed offsets are considerably shorter than the characteristic QPP periods of 1–7 minutes, supporting the close temporal correspondence of the oscillations across the EUV, UV, and HXR wavebands. The near-simultaneous detection of these oscillations by three independent instruments provides strong evidence that the QPPs are associated with the solar flare.

\section{The Non-Force-Free-Field Extrapolation of Active Region 13842}
\label{exp-mod}

To investigate the key magnetic features underlying the onset of the X9.0 class flare in AR 13842, we construct the coronal magnetic field line configuration using the NFFF extrapolation model developed by \citet{2008ApJ...679..848H} and \citet{2010JASTP..72..219H}. The model is based on the principle of Minimum Dissipation Rate (MDR) under which the plasma relaxes towards a state that minimizes the rate of energy dissipation. The magnetic field is then represented as a superposition of a potential field and two linear force-free fields. For a detailed description of the extrapolation model, we refer the readers to \citet{2008SoPh..247...87H,2008ApJ...679..848H,2010JASTP..72..219H}.

In order to perform the NFFF extrapolation, we use the photospheric vector magnetic fields of Helioseismic and Magnetic Imager \citep[HMI,][]{2012SoPh..275..229S} onboard SDO. In particular, we consider the \texttt{hmi.sharp\_cea\_720s} data series with $0.5''$ per pixel resolution and a time cadence of 12 minutes. The magnetogram is initially remapped onto the Lambert cylindrical equal area (CEA) projection, and then it is transformed into a heliographic coordinate \citep{1990SoPh..126...21G} to obtain the magnetic field in a Cartesian grid. To explore the change in magnetic configuration during the flaring event, we have performed a series of extrapolations of the AR with the aforementioned data series for the time interval from 11:00 UT to 12:24 UT, covering the whole flaring period with a time cadence of 12 minutes. To focus only on the flaring region, we have cropped the original cutout to $850 \times 450$ grid points along the $x$ and $y$-directions, respectively. While $450$ grid points are taken along $z$-direction. The physical domain corresponds to $\approx $308 Mm in $x$ and $\approx $163 Mm in both $y$ and $z$-directions. It is important to note that the vector magnetograms used for the NFFF extrapolations at 12:12 UT and 12:24 UT were acquired during the flare's evolution, including the impulsive phase. Strong flare emission can affect the inversion of the Zeeman profiles and may result in abnormal magnetic-field values or bad pixels in HMI vector magnetograms \citep{Hoeksema_2014}. To assess whether such effects could influence our extrapolations, we carefully inspected the corresponding photospheric vector magnetograms. We found no conspicuous bad pixels, anomalous field values, or other evident artifacts associated with the flare emission. In addition, the magnetic-field values remained within the HMI dynamic-range limits reported by \citet{Hoeksema_2014}. These checks indicate that the vector magnetograms used in our analysis were not significantly affected by flare-related measurement artifacts and therefore provide a reliable lower boundary for the NFFF extrapolations.  

It is noteworthy that although NLFFF extrapolations are more widely used for reconstructing non-linear coronal magnetic fields, previous studies have demonstrated that NFFF extrapolations can successfully reproduce complex non-potential magnetic configurations, including twisted flux ropes, sheared arcades, QSLs, and HFTs, and can provide useful insights into the magnetic topology associated with the initiation and evolution of various solar transient events \citep{2020ApJ...903..129P,2022FrASS...9.2108B,2023A&A...677A..43P,2025SoPh..300...22K,2026SoPh..301...67K}. These studies demonstrate that, despite its simplifying assumptions, the NFFF approximation can provide a useful description of the 3D non-potential magnetic structures in ARs. In the present study, we therefore employ NFFF to investigate the flare-associated magnetic configuration.

\section{Results}
\label{result}

\subsection{Initial extrapolated magnetic configuration}
\begin{figure}
   	\centering
	\includegraphics[width=1.0\linewidth]{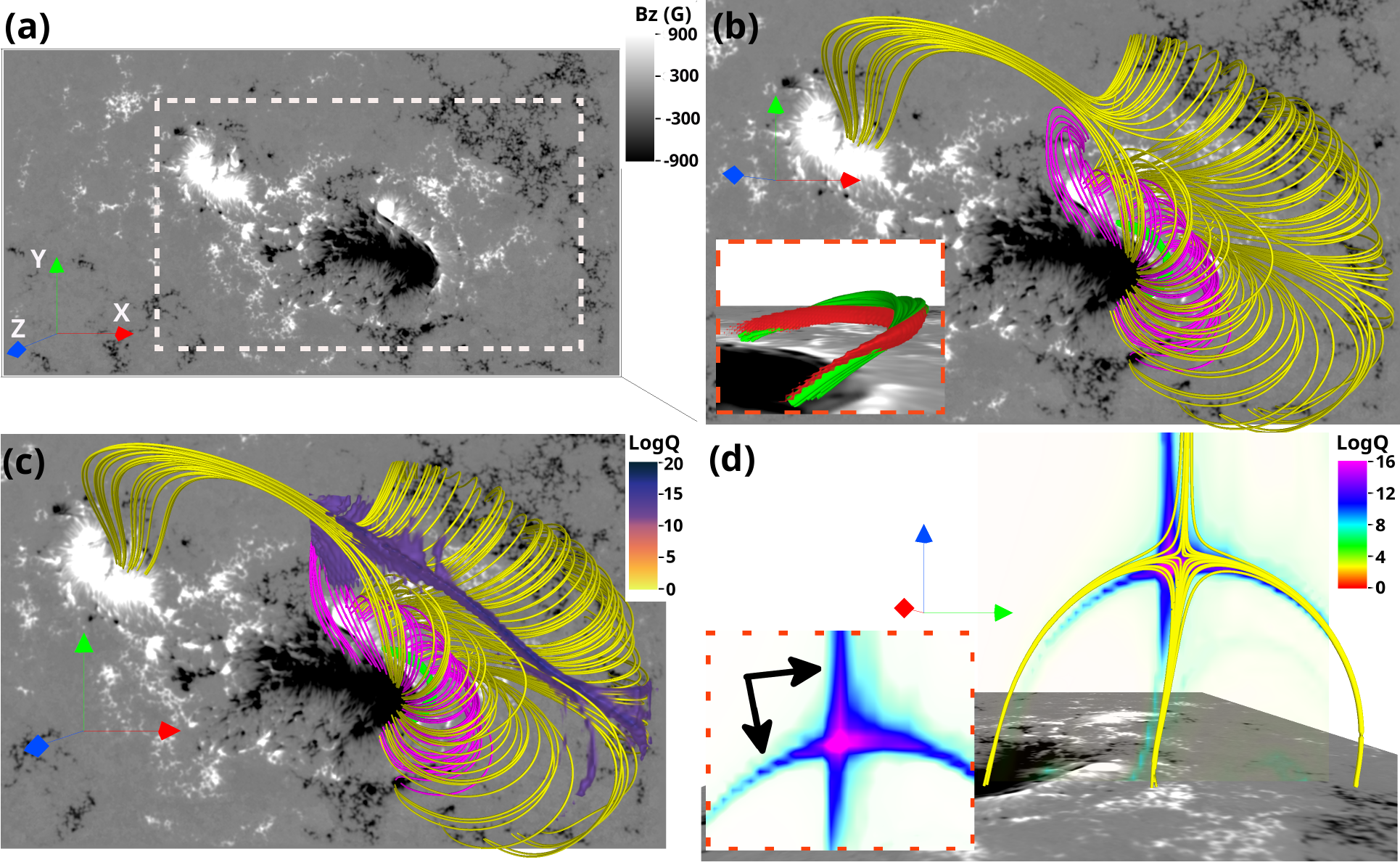}
	\caption{Extrapolated magnetic configuration at 11:00 UT. Panel (a) shows the $B_z$ component on the bottom boundary with dashed box enclosing the flaring region. Panel (b) shows the top view of extrapolated field lines, where yellow lines construct the HFT structure having a fan-spine structure, pink lines represent the sheared arcade and green lines correspond to the flux rope. The inset in panel (b) highlights the flux rope with isosurface of $T_w \approx 1.33$.  Panel (c) is further overlaid with the $LogQ \geq 8$ distribution near the HFT showing high gradient in field line connectivity along the HFT fan. In panel (d), a 2D projection of the yellow field lines along with the squashing factor is plotted on a $x$-constant plane. The figure inset highlights the intersection of two QSLs (black arrows) in the vicinity of the X-shaped projected field lines. }
     \label{exp_1100h}
\end{figure}

\vspace{0.2 cm}
Figure \ref{exp_1100h} illustrates the extrapolated magnetic configuration at 11:00 UT, which corresponds to the pre-flare stage. Panel (a) shows the Bz component of the magnetogram at the bottom boundary ($z=0$) of the computational domain. The white dashed box encloses the flaring region where the flare brightenings are observed. Panel (b) depicts the extrapolated field lines in the flaring region. Noticeably, in this and subsequent figures, we have utilized VAPOR software \citep{2019Atmos..10..488L} for the visualization of magnetic field lines and related parameters. Similar to a 3D magnetic null configuration, the yellow colored field lines appear to constitute a dome-like fan structure and anchored spine rooted in a major positive polarity region. However, we have verified that there is a complete absence of nulls in the flaring region  using the trilinear method of null detection \citep{2007PhPl...14h2107H,2020A&A...644A.150O}. This suggests that the yellow colored field lines correspond to an HFT structure \citep{2026SoPh..301...67K}. The pink field lines located under the HFT represent a sheared arcade. The field lines in green underneath the sheared arcade seem to represent a flux rope. For better legibility, in the inset of panel (b), the green field lines are further overlaid with the isosurface of twist parameter ($T_{w}$) with isovalue $\approx$ 1.33 (in red color). $T_{w}$ is calculated using the equation,
\begin{equation}
    T_{w}={\frac{1}{4 \pi} \int_l {\alpha} dl} 
\end{equation}
\noindent where {$\alpha$ = $\mathbf{\frac{J.B}{|B|^{\text{2}}}}$} and $dl$ is a line element \citep{2006JPhA...39.8321B}. 
Clearly, the green field lines with $T_w \approx 1.33$ constitute a flux rope. The identification of the flux rope in our NFFF extrapolation can be compared with previous investigations of the same X9.0 flare. \citet{2025ApJ...985L..16D} identified three flux rope structures using NLFFF extrapolations. In contrast, \citet{2025ApJ...991...77M} employed an NLFFF derived magnetic field as the initial condition for a data-constrained MHD simulation and found a flux rope embedded within the surrounding sheared field. Our NFFF extrapolation similarly reveals one flux rope embedded within a sheared arcade, together with an HFT in the flaring region. The difference between the three flux rope configuration inferred by \citet{2025ApJ...985L..16D} and the single flux rope obtained by \citet{2025ApJ...991...77M} and in the present study suggests that the detailed fine-scale magnetic topology may depend on the extrapolation and modeling approach. Nevertheless, the common identification of a strongly non-potential, twisted magnetic structure embedded within a sheared magnetic environment supports the presence of a highly stressed core field in the flaring region.

To further confirm the presence of the HFT, Figure \ref{exp_1100h}(c) plots the extrapolated field lines superimposed with the squashing factor \citep[Q,][]{2002JGRA..107.1164T} on a logarithmic scale, which measures the mapping of the magnetic field lines. The Q is calculated using the code by \citet{2016ApJ...818..148L}. Important are the high values of $LogQ$ ($\geq8$) along the dome-like fan structure --- suggesting the high gradient in the field line connectivity in this region. To delve deeper, Figure \ref{exp_1100h}(d) depicts $LogQ$ in a $x$-constant plane superimposed with the field lines of the transverse field (achieved by setting $B_x=0$ in $\mathbf{B}$ for the yellow field lines), which signify the projected magnetic field lines on the $x$-constant plane. The intersection of the two QSLs is clearly observable (inset of panel (d)), along with the existence of X-shaped field lines of the transverse field near the QSLs. Such convergence of QSLs alongside the X-shaped configuration of the projected field lines can be recognized as HFT \citep{2002JGRA..107.1164T, 2021SoPh..296...26K,2026SoPh..301...67K}.

%-------

\subsection{Evolution of the magnetic configuration through the series of extrapolations}
\vspace{0.2 cm}

\begin{figure}
   	\centering
	\includegraphics[width=1\linewidth]{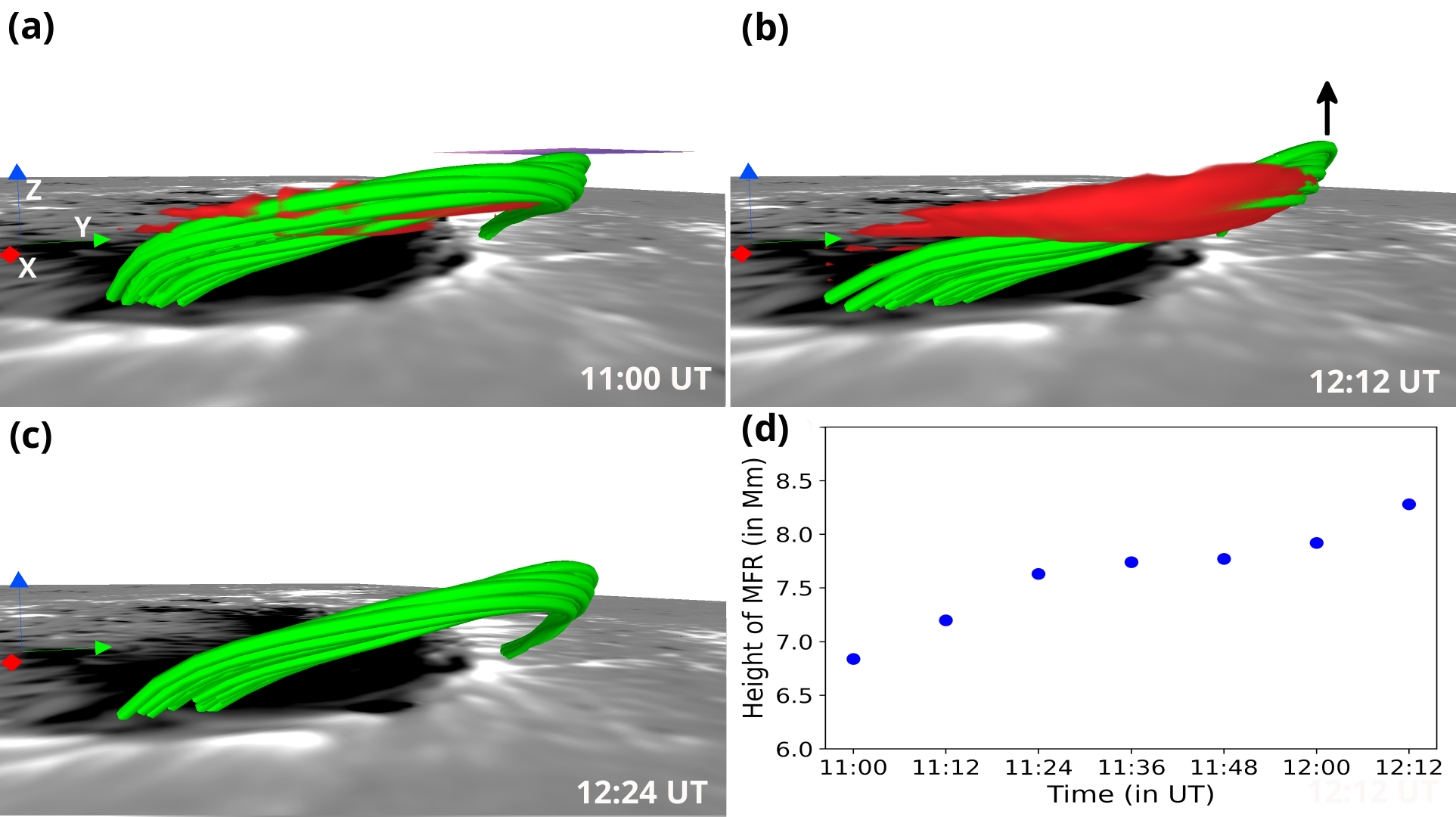}
	\caption{Panels (a)-(c) show the flux rope overlaid with $T_{w}$ having values greater than one (in red color). The values are found to increases from $\approx$ 1.33 to $\approx$ 1.43 for a time period ranging from 11:00 UT to 12:12 UT (panels (a) and (b)). The plane in panel (a) is considered to track the rise of the rope. The rise is marked by a black arrow in panel (b). Panel (c) shows the disappearance of the high twist values at 12:24 UT after the flare peak ($\approx$ 12:18 UT). Panel (d) plots the height of the flux rope from 11:00 UT to 12:12 UT, estimated from the series of the extrapolations. }
     \label{MFR_TW_DI}
\end{figure}

To have an insight of the evolution of the key magnetic features during the flare, we utilize the series of the NFFF extrapolation starting from 11:00 UT with a time resolution of 12 minutes. In Figure \ref{MFR_TW_DI}, we focus on the evolution of the flux rope. Panels (a)-(c) plot the field lines of the rope overlaid with $T_{w}>1$ in red color.  To identify and visualize the field lines associated with the flux rope, seed points were initially selected from the region of enhanced \(T_w\) at 11:00 UT. These seed points were then biased using the \(T_w\) variable in VAPOR software, such that the traced field lines remained concentrated within regions of elevated $T_w$ throughout the extrapolation sequence. The same $T_w$-based selection criterion was applied at each extrapolation time to minimize the dependence of the identified flux rope structure on the choice of seed points. The height of the flux rope at each extrapolation time was estimated by placing a horizontal plane at the uppermost extent of the selected flux rope field lines, parallel to the $z=0$ plane of the computational domain (see Figure \ref{MFR_TW_DI}(a)). The vertical distance of this plane from the $z=0$ plane was then taken as the estimated height of the flux rope. 
Importantly, for a time interval from 11:00 UT to 12:12 UT, maximum $T_{w}$ of the rope increases from $\approx$ 1.33 to $\approx$ 1.43 (panels (a)-(b)). However, after the flare peak ($\approx$ 12:18 UT), the high values of $T_{w}$ disappear and it becomes lower than one (which is masked during plotting) as illustrated in panel (c). This indicates the absence of the flux rope after the flare peak. Moreover, the flux rope shows an apparent rise in the vertical direction, marked by the black arrow in panel (b). Panel (d) plots the height of the flux rope at different times during the pre-flaring and the initiation stages, which clearly documents the ascent of the rope. 

\begin{figure}
    \centering
    \includegraphics[width=1.0\linewidth]{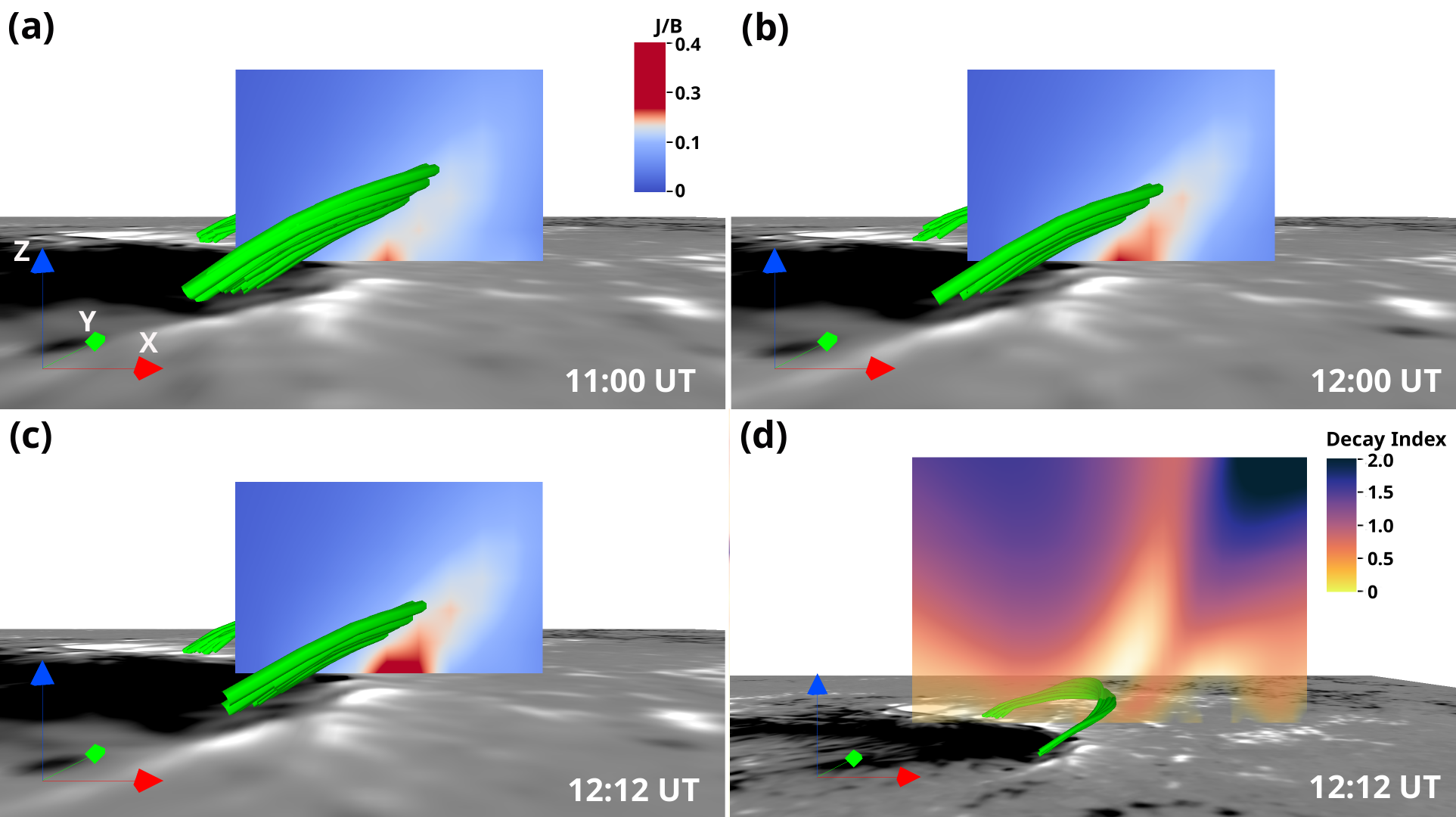}
    \caption{Panels (a) to (c) plot the flux rope overlaid with $|\textbf{J}|/|\textbf{B}|$ at a $y$-constant plane at different extrapolation times keeping $B_z$ at the bottom boundary. Notable is an increase in $|\textbf{J}|/|\textbf{B}|$ as time progresses toward to flare peak, signaling to onset of reconnection in sheared arcade. In panel (d), the rope is overlaid with the decay index ($n$) at 12:12 UT. The torus unstable region with $n \approx 1.5$ is located above the rope.}
    \label{MFR_JbyB}
\end{figure}

To explore the possible physical mechanism underlying the rise of the flux rope, in Figures \ref{MFR_JbyB}(a)-(c), we plot the flux rope overlaid with $|\textbf{J}|/|\textbf{B}|$. Important is the notable enhancement in $|\textbf{J}|/|\textbf{B}|$ beneath the flux rope as time advances towards the flare onset time (panels (a) and (b)). Subsequently, we observe a more rapid increase in $|\textbf{J}|/|\textbf{B}|$ from 12:00 UT to 12:12 UT (panel (c)). Notably, such an enhancement in $|\textbf{J}|/|\textbf{B}|$ may correspond to an increase in the magnetic field gradient of the field lines of the sheared arcade, as the foot-points of the field lines rooted in the opposite polarities move in closer proximity \citep{2016ApJ...828...62J,2018NatCo...9..174I}. As a result, the high values of $|\textbf{J}|/|\textbf{B}|$ below the rope can ultimately lead to the onset of magnetic reconnection in the sheared arcade which may contribute to the twist increase of the flux rope and its ascent {\citep{2016ApJ...830...80K, Kumar2016Continuous, 2023A&A...677A..43P}}.

Next, to explore the role of the overlying strapping field on the flux rope in relation to the torus instability, in Figure \ref{MFR_JbyB}(d), we illustrate the flux rope overplotted with the decay index (n), expressed by 
$ n=-{\frac{d \ln B_{ex}}{d \ln h}}$, where $B_{ex}$ is the horizontal component of the external magnetic field that surrounds the flux rope at height $h$. The decay index is a measure of rate of decay of horizontal field with height and a critical value of $n\approx 1.5$ indicates the rope to be torus unstable \citep{2006PhRvL..96y5002K}. It is noteworthy that a torus unstable region exists above the flux rope, suggesting the potential for the ascending rope to become torus unstable upon reaching that area under suitable conditions \citep{2013ApJ...771L..30J,2018NatCo...9..174I}. Nevertheless, capturing the dynamic ascent of the rope is not feasible through a series of extrapolations and requires MHD simulation. 
%This statement is in line with previous study of \citet{2025ApJ...985L..16D}.% 

\begin{figure}
   	\centering
	\includegraphics[width=1\linewidth]{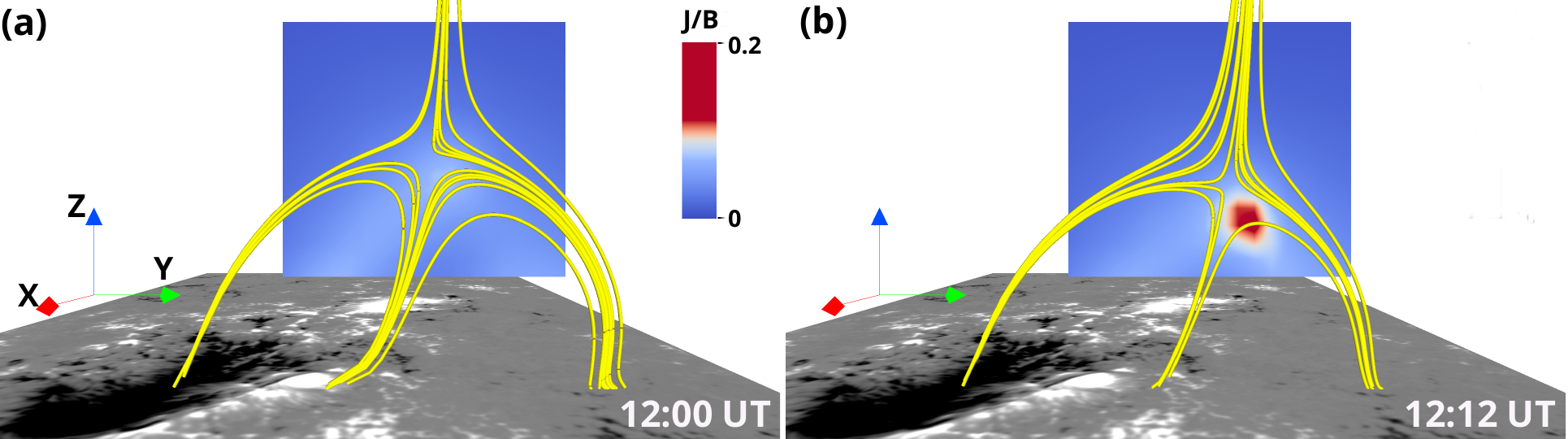}
	\caption{Plots of the projected field lines of the HFT at a $x$-constant plane at 12:00 UT (panel (a)) and 12:12 UT (panel (b)). The plots are also overlaid with $|\textbf{J}|/|\textbf{B}|$. Notable is an increase in $|\textbf{J}|/|\textbf{B}|$ in the vicinity of the X-shaped geometry. }
    \label{hft-rec}
\end{figure}

To investigate the potential of magnetic reconnection at the HFT, in Figure \ref{hft-rec}, we present the field lines of the transverse field near the HFT (as illustrated in Figure \ref{exp_1100h}(d)) at 12:00 UT and 12:12 UT. Additionally, the figure is overlaid with  $|\textbf{J}|/|\textbf{B}|$. It is noteworthy that there is an increase in $|\textbf{J}|/|\textbf{B}|$ near the X-shaped geometry at 12:12 UT when compared to its values at 12:00 UT. This increase is associated with a rise in the gradient of the magnetic field in the vicinity of the HFT --- advocating the onset of magnetic reconnection at the HFT \citep{2026SoPh..301...67K}. Importantly, the reconnection occurring at the HFT may also plays a role in eliminating the overlying flux of the flux rope and the sheared arcade, which favors the eruption of the flux rope.

\subsection{Comparison of the extrapolated configuration with the observations}
\label{exp-obs}
\vspace{0.2 cm}

\begin{figure}
    \centering
    \includegraphics[width=1.0\linewidth]{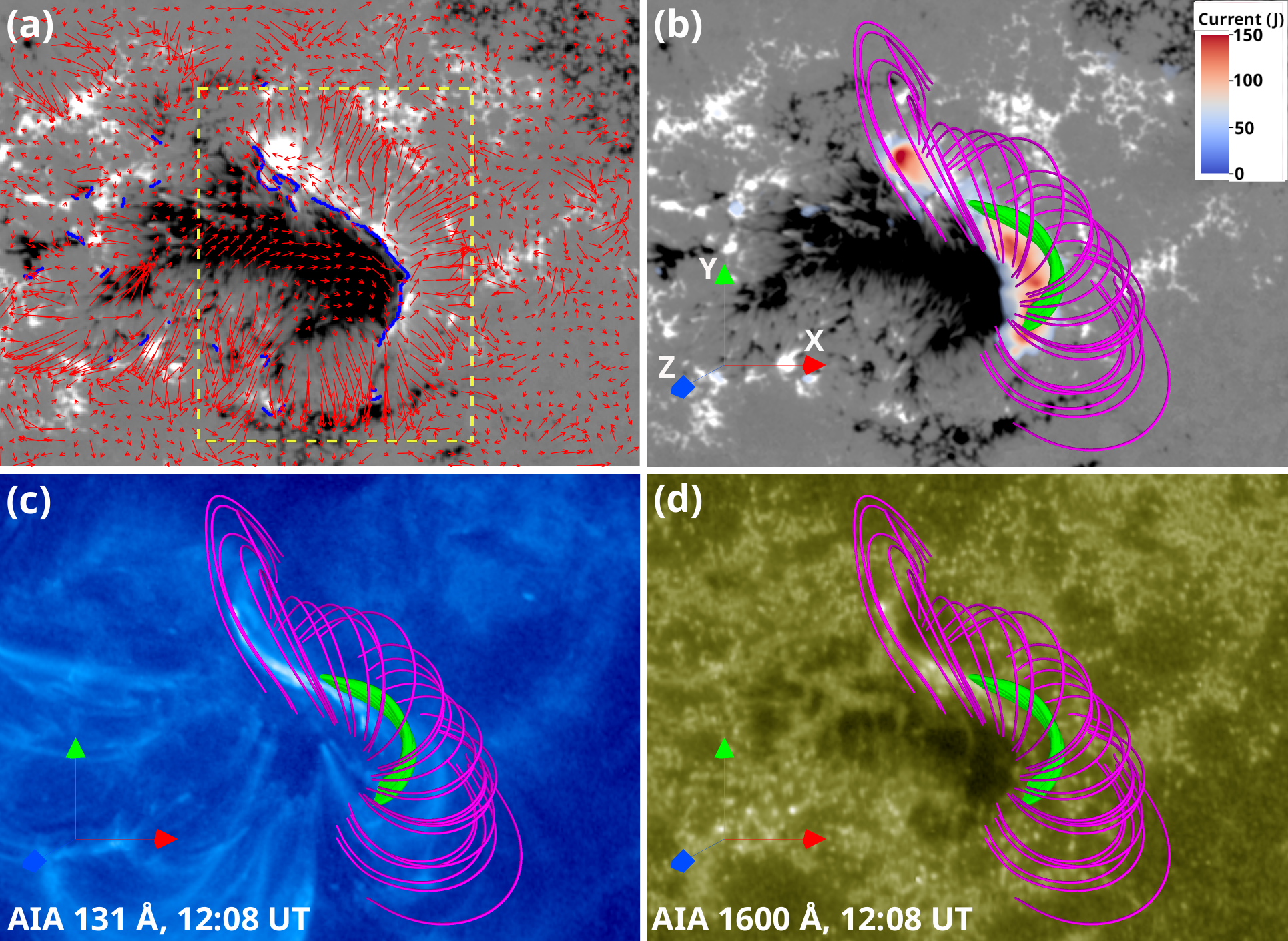}
    \caption{ Panel (a) plots the horizontal velocity field (red arrows) of photosphere using $\text{DAVE4VM}$ method, and superimposed on Bz at 12:00 UT along with the PILs (blue curves). The panel illustrates the oppositely directed plasma flows across the extended PIL (within yellow rectangle) can potentially induce shear in arcade field lines. In panel (b) sheared arcade is overlaid with Bz at 12:00 UT. The panel is further superimposed with $\mathbf{|J|}$ (in the units of $2.21 \times 10^{-4}\, \text{Am}^{-2}$)} at the bottom boundary, which documents the presence of the high current near the sheared arcade at 12:00 UT. Panels (c) and (d) show the sheared arcade (at 12:00 UT) overlaid with AIA 131 and 1600 channels (at 12:08 UT) respectively. Notable is the similarity of the sheared arcade with the brightenings in the different AIA channels. All the panels are overlaid with the flux rope (in green).
     \label{sigmoid}
\end{figure}

Here we compare the sequential dynamics derived from the series of extrapolations with the different channels of AIA. Given the constraints of time resolution in the series of extrapolations, we have selected AIA images based on the development of flare brightenings/ribbons and compared them with the proximate timestamps of the extrapolated field lines. To investigate the pre-flare evolution of this event, Figures \ref{sigmoid}(a) show the horizontal photospheric velocity field (red arrows) derived from the vector magnetograms at 11:48 UT and 12:00 UT using the Differential Affine Velocity Estimator for Vector Magnetograms (DAVE4VM) technique \citep{2008ApJ...683.1134S}. The derived velocity field is overlaid on the $B_z$ map at 12:00 UT, with the polarity inversion lines (PILs) indicated by blue curves. We find that the plasma flows across the core PIL (in yellow rectangle) are directed oppositely on either side of the PIL, consistent with the velocity pattern reported by \citet{2025ApJ...985L..16D}. Such oppositely directed motions are expected to progressively enhance the magnetic shear of the overlying arcade \citep{2022FrASS...9.2108B}, thereby increasing the electric current density (|$\mathbf{J}$|) within the arcade, as shown in Figure \ref{sigmoid}(b). Notably, the field lines of the sheared arcade shown in Figures \ref{sigmoid}(c) and (d) closely resemble the S-shaped brightening BR1 observed in AIA 131~{\AA} (Figure \ref{131_obs}(b)) and the faint brightening FB seen in AIA 1600~{\AA} (Figure \ref{1600_obs}(a)), both of which appeared at the onset of the flare. The spatial correspondence between these brightenings and the highly sheared, current-carrying arcade suggest that the enhanced electric currents in the sheared arcade may be closely associated with the formation of BR1 and FB \citep{2020ApJ...903..129P}.

\begin{figure}
    \centering
    \includegraphics[width=1.0\linewidth]{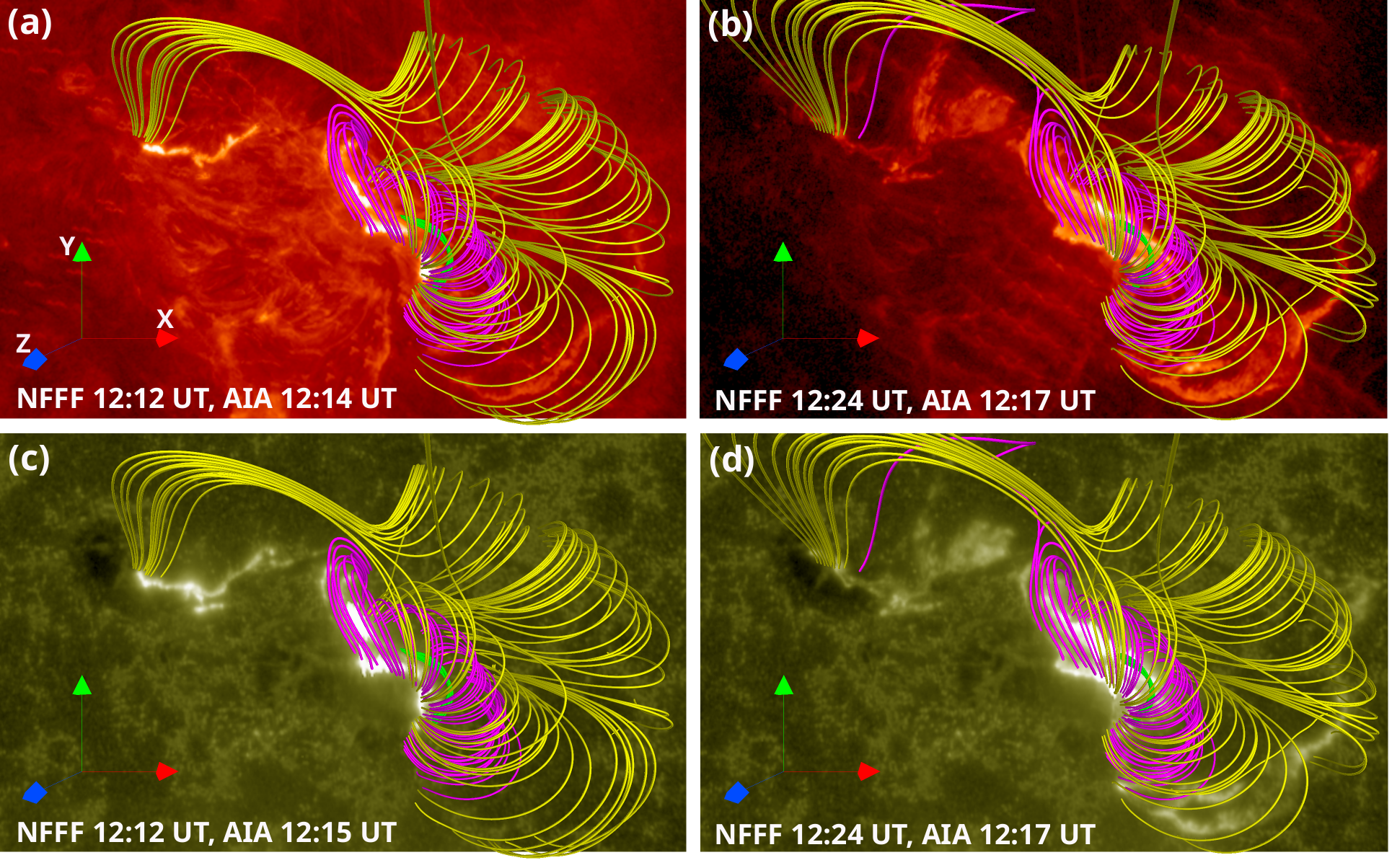}
    \caption{The extrapolated field lines are overplotted with the AIA 304 {\AA} (panels (a) and (b)) and 1600 {\AA} ((panels (c) and (d)) images. The figure correlates the reconnection in sheared arcade and the HFT with all brightenings and ribbons.  }
     \label{aia-exp}
\end{figure}

Figure \ref{aia-exp} illustrates the extrapolated field lines superimposed with AIA 304 {\AA} (panels (a) and (b)) and 1600 {\AA} (panels (c) and (d)) images at times close to the flare peak. Notably, the foot-points of the sheared arcade are nearly co-spatial with the parallel brightenings BR1 and BR2 (Figure \ref{131_obs} and Figure \ref{aia-exp}(a)) as well as the parallel flare ribbons R1 and R2 (Figure \ref{1600_obs} and Figure \ref{aia-exp}(c)). The continuous foot-point motion of the sheared arcade (Figure \ref{sigmoid}(a)) following the flare onset may facilitate the approach of non-parallel field lines, resulting in an increase in $|\textbf{J}|/|\textbf{B}|$ beneath the flux rope (Figure \ref{MFR_JbyB}(c)). This increase in $|\textbf{J}|/|\textbf{B}|$ could ultimately lead to magnetic reconnection occurring within the sheared arcade located beneath the flux rope. The charged particles accelerated from this reconnection may contribute to the formation of the parallel brightenings BR1 and BR2 (Figure \ref{131_obs} and Figure \ref{aia-exp}(a)) as well as the parallel flare ribbons R1 and R2 (Figure \ref{1600_obs} and Figure \ref{aia-exp}(c)). This phenomenon aligns with the CSHKP model \citep{1964NASSP..50..451C,1966Natur.211..695S,1974SoPh...34..323H,1976SoPh...50...85K}. Furthermore, in Figures \ref{aia-exp}(a) and (c), the foot-points of the anchored HFT spine closely follow the remote brightening BR3 (Figure \ref{131_obs}) and the ribbon R3 (Figure \ref{1600_obs}). The charged particles accelerated by the HFT reconnection (Figure \ref{hft-rec}) may travel along the anchored HFT spine, potentially causing the brightening BR3 and the ribbon R3. As time advances towards the peak of the flare, as shown in Figures \ref{aia-exp}(b) and (d), we observe that the foot-points of the HFT fan are nearly co-spatial with the quasi-circular brightening BR4 (Figure \ref{131_obs}) and the ribbon R4 (Figure \ref{1600_obs}). The accelerated charged particles through the HFT reconnection are also anticipated to travel along the HFT fan, which may result in the quasi-circular brightening BR4 and the ribbon R4.

As discussed above, the extrapolation sequence indicates the recurrence of magnetic reconnection within the sheared arcade and the HFT in the flaring region. In addition, \citet{2026A&A...710A.105P} reconstructed the HXR emission in the 14–18 keV energy range for the same event and found that this energy range is dominated by thermal emission, with a significant non-thermal contribution also present. In this context, a plausible scenario is that recurrent magnetic reconnection episodes repeatedly energize the plasma in the solar atmosphere, contributing to the observed HXR emission and giving rise to the QPPs during the impulsive phase of the flare \citep{ZIMOVETS201817,Clarke_2021,2025CosRe..63..476Z}. Similar interpretations of QPPs in terms of recurrent or repetitive magnetic reconnection have also been discussed in previous studies of two-ribbon flares, including events with a flux rope embedded in a sheared magnetic arcade \citep{ZIMOVETS201817,2025CosRe..63..476Z}. This proposed mechanism is also consistent with the recent study of \citet{2025JGRA..13033772L}. However, this interpretation remains speculative, as the present study does not include time-dependent MHD simulations capable of directly validating the proposed reconnection dynamics.

The different QPP periods observed in the HXR, UV, and EUV wavebands may arise from the different physical responses of the solar atmosphere to repetitive energy release and particle acceleration during magnetic reconnection. HXR emission provides a relatively prompt signature of accelerated electrons precipitating into the chromosphere and can therefore respond rapidly to individual reconnection episodes. In contrast, the EUV/UV emissions observed by AIA are largely influenced by the subsequent heating, cooling, and plasma evolution in the transition region and corona, and hence may respond differently to the same sequence of energy-release episodes \citep{1985ApJ...289..434F,2009ApJ...699..968M,2012SoPh..275...17L}. Consequently, a periodicity that is prominent in HXR may not necessarily appear with the same strength or detectability in EUV/UV emission, and vice versa. In addition, differences in instrumental cadence, temporal resolution, photon-counting statistics, and signal-to-noise ratio can influence the detectability of particular periodicities in different wavebands. Similar wavelength-dependent QPP periods have also been reported in previous multi-wavelength studies \citep{2015A&A...574A..53K,2025JGRA..13033772L}.

\section{Summary and Discussion}
\label{Summary} 
In this work, we examine the X9.0 class flare and associated QPPs in NOAA AR 13842 on 2024 October 3. For this purpose, we employ multi-wavelength observations from AIA, HMI/SDO, HEL1OS/Aditya-L1 and STIX/SO. Additionally, we supplement these observations with a series of NFFF extrapolations (covering the whole flaring duration) to gain insights into the overall flaring process. 
From the multi-wavelength observations, we identify four distinct flare brightenings and ribbons, which differ in shape and are designated as BR1-BR4 (Figures \ref{131_obs} and \ref{304_obs}) and R1-R4 in (Figure \ref{1600_obs}). Notably, oscillations are detected in the detrended EUV, UV, and HXR light curves, displaying three distinct peaks during the impulsive phase of the flare, along with smaller peaks during the decay phase of the flare, as illustrated in Figures \ref{qpp_aia} and \ref{qpp_hel1os}, respectively. Following a subsequent wavelet analysis, these oscillations are characterized as QPPs with multiple periods $\approx$ 2 minutes and $\approx$ 6 minutes in EUV; $\approx$ 7 minutes in UV; and $\approx$ 1 minute in HXR.

The extrapolated magnetic configuration of the flaring region at the pre-flaring stage reveals the presence of key magnetic structures such as a low-lying flux rope, a sheared arcade above the flux rope and an HFT enveloping the rope and the sheared arcade (Figure \ref{exp_1100h}). The series of the extrapolation documents a continuous elevation of the twist of the rope along with its ascent towards a torus unstable region from the pre-flaring stage to the peak of the flare. Important is the enhancement of $|\textbf{J}|/|\textbf{B}|$ below the flux rope (Figure \ref{MFR_JbyB}) --- a tell-tale sign of the onset of magnetic reconnection inside the field lines of the sheared arcade. Such reconnections, when repeated over time, may ascend the flux rope and result in its eventual destabilization \citep{2016ApJ...830...80K}. Importantly, the sequential evolution from the series of the extrapolations also provides evidence of magnetic reconnection at the HFT (Figure \ref{hft-rec}). This reconnection might have also played a role in diminishing the magnetic confinement above, thereby promoting the destabilization and possible eruption of the flux rope. 

The horizontal field velocity derived from the DAVE4VM technique supports the gradual motion of the foot-points of sheared arcades
%The foot-points of the sheared arcade show a gradual motion 
that is favorable for the increment of the shear and electric current associated to the arcade. This increased current appears to cause the plasma heating and leads to S-shaped sigmoidal brightening (Figure \ref{sigmoid}). With time, the motion can also facilitate non-parallel field lines to come in close proximity. This can lead to the enhancement in $|\textbf{J}|/|\textbf{B}|$ below the flux rope and consequent magnetic reconnections inside the sheared arcade. As the foot-points of the sheared arcade are almost coinciding with the parallel flare brightenings and ribbons, these reconnection are potentially involved in the generation of the brightenings and the ribbons. Furthermore, the foot-points of the anchored spine and fan of the HFT trace the remote and the quasi-circular brightenings as well as the ribbons --- indicating the role of reconnection at the HFT in the development of these brightenings and ribbons.

Notably, \citet{2025A&A...695L...4L} investigated QPPs in the same flare using high-time-cadence observations and reported short-period QPPs of $\approx$1 s in the higher energy ($>20$ keV) HXR and $\gamma$-ray channels. Their Fourier power spectra also show enhanced power at periods longer than $\sim$10 seconds, although these longer-period signatures were not explored in detail because their analysis primarily focused on the shorter period QPPs. In contrast, our study focuses on the lower energy HXR range (11--20 keV) and identifies a QPP with a period of approximately 1 minute, while also revealing longer period QPPs in the UV and EUV wavebands. Thus, the present study complements the results of \citet{2025A&A...695L...4L} by extending the investigation from the previously reported short period QPPs to longer period oscillations and their multi-wavelength characteristics. Interestingly, as discussed above, the repeated reconnection episodes that govern the various observational signatures of the flaring event may also contribute to the multi-periodic behavior of QPPs observed across different wavelengths. The intermittent nature of these reconnection processes can sustain ongoing energy release, leading to repeated particle acceleration and plasma heating, which may produce quasi-periodic signatures in the HXR emission and corresponding thermal responses in the UV and EUV channels. This interpretation is consistent with earlier studies \citep{2021ApJ...910..123C,2025JGRA..13033772L,2026NatAs..10...13L,2026NatAs..10...54A}.

Overall, this study attempt to identify the role of recurrent magnetic reconnection inside the sheared arcade and the HFT in giving rise to the X-class flare and the associated QPPs. The presence of the ascending flux rope towards the torus unstable region potentially explains the eruptive nature of the flare.
It is important to note that \citet{2025ApJ...985L..16D} has proposed the existence of a system of three flux ropes, which are believed to be formed through a collisional shearing process, located near the polarity inversion line. The eruption of this system is thought to trigger the flare. \citet{2025ApJ...991...77M} indicated that the tether-cutting reconnection occurring within the sheared field lines is responsible for the formation of the flux ropes, with the eruption being driven by torus instability and magnetic reconnection. The formation of the remote and quasi-circular brightenings in this event can be further compared with the interpretation of \citet{2025ApJ...991...77M}. In \citet{2025ApJ...991...77M}, the remote brightening was interpreted as resulting from reconnection between the foot-points of the flux rope and adjacent coronal loops, while the quasi-circular brightenings, identified as two distinct brightenings, were attributed to reconnection between the flux rope and overlying loops. In comparison, our analysis identifies two additional parallel brightenings in the central flaring region that are associated with tether-cutting type reconnection within the sheared arcade. Such reconnection can contribute to the build-up and rise of the flux rope and is therefore broadly consistent with the eruption mechanism proposed by \citet{2025ApJ...991...77M} based on their MHD simulations. Furthermore, we also propose that reconnection at the HFT may contribute to the formation of the remote and quasi-circular brightenings observed in our study. Thus, the two studies are broadly consistent in associating the observed brightenings and flux rope evolution with magnetic reconnection, while our analysis provides an additional magnetic topological perspective in which HFT reconnection may link the evolution of the coronal magnetic structure to the formation of the remote and quasi-circular brightenings. Our analysis further develops the investigation of the QPPs linked to the flare by uncovering the presence of significantly longer periodicities ($\approx$ 1 minute) in lower energy range (11-20 keV) of HXR channel. Additionally, the present study reports the long period QPPs in UV and EUV wavebands. Nevertheless, without time-dependent MHD simulations, the temporal evolution of the reconnection process cannot be established, and therefore the role of recurrent magnetic reconnection in QPPs cannot be confirmed conclusively. In the future, we intend to tackle this issue using an advanced modeling approach that includes the MHD simulations.

\appendix
\section{Appendix}\label{appendix1}

\begin{figure}
    \centering
    \includegraphics[width=0.9\linewidth]{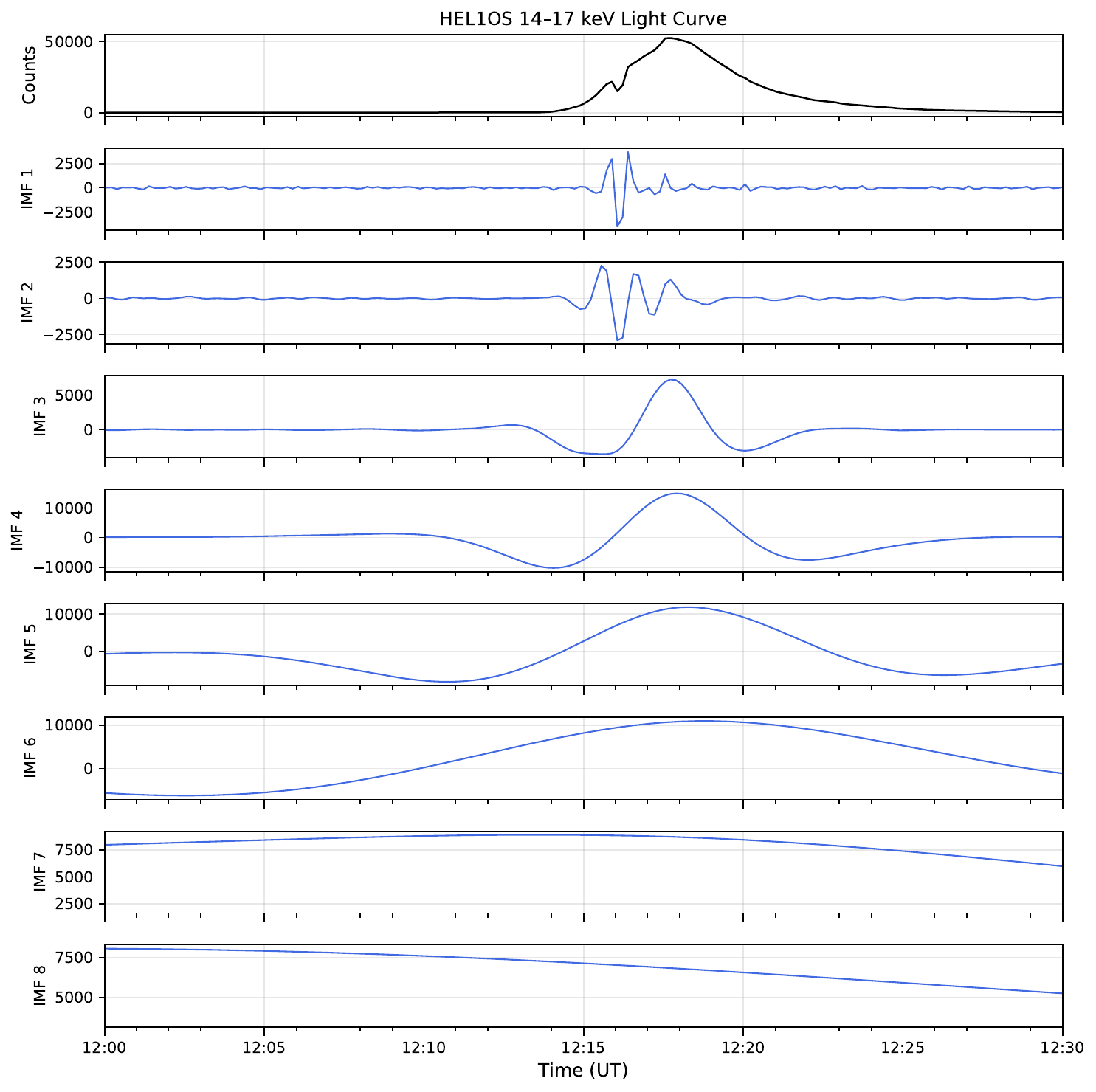}
    \caption{EEMD of the HELIOS HXR light curve in 14–17 keV energy band. The original light curve is shown in the top panel, followed by the extracted IMFs (IMF-1 to IMF-8). }
    \label{IMFs}
\end{figure}

The HXR light curves are decomposed into IMFs using EEMD. The resulting decomposition is shown in Figure \ref{IMFs} (for 14--17 keV). The extracted IMFs represent oscillatory modes with progressively decreasing characteristic frequencies, while the higher order IMFs describe the slowly varying flare profile and background trend. To identify the IMF as a QPP candidate, each extracted mode was independently investigated using Fourier power spectrum and Morlet wavelet analyses. Only IMF-2 exhibits a coherent oscillatory signal (with three consecutive peaks) confined to the impulsive phase of the flare, and its dominant period agrees with the statistically significant period detected from the Fourier and Morlet wavelet analyses. In contrast, IMF-1 is characterized primarily by high-frequency fluctuations or can be a instrument noise that do not exhibit a stable periodic signature, while IMF-3 and higher-order IMFs represent lower-frequency variations associated with the overall flare evolution rather than the QPP so they are considered as the background trend. Based on the consistency between the EEMD decomposition and the independent Fourier and Morlet wavelet analyses, IMF-2 is treated as the mode containing the QPP signal and is used for all subsequent analyses presented in this work.

\section*{Acknowledgments}
We acknowledge the use of the visualization software VAPOR 
(\url{www.vapor.ucar.edu}) for generating relevant graphics. The computations were performed on high performance computer of Department of Physics, Patna University, Patna, India. Data and images are courtesy of NASA/SDO and ISRO/Aditya L1 and the HMI, AIA and HEL1OS science teams. SDO/HMI as well as Aditya L1/HEL1OS are joint efforts of many teams and individuals to whom we are greatly indebted for providing the data. We also acknowledge the science team of Solar Orbiter which is a space mission of international collaboration between ESA and NASA, operated by ESA. The STIX instrument is an international collaboration between Switzerland, Poland, France, Czech Republic, Germany, Austria, Ireland, and Italy. Part of the computation was carried out on the computing cluster Pegasus of IUCAA, Pune, India. S.K., S. and R.C would like to acknowledge the support from the ANRF-SERB project No. SUR/2022/000569. S.K. also acknowledges support from Patna University project No. RDC/MRP/07. S.S.N. acknowledges the supports from the I+D+I project PID2023-147708NB-I00 funded by FEDER,EU and MICIU/AEI/10.13039/501100011033/. We thank the reviewers for providing insightful suggestions and comments which led to the significant improvement of this paper.

\section*{Data availability}
The data that support the findings of this study are available from the corresponding author upon reasonable request.

\newpage

\bibliographystyle{cas-model2-names}
\bibliography{ref}

\end{document}